
\input harvmac
\def\CS{{\cal S}}

\def\im{{\rm Im}}
\def\re{{\rm Re}}
\font\cmss=cmss10 \font\cmsss=cmss10 at 7pt
\def\IZ{\relax\ifmmode\lrefhchoice
{\hbox{\cmss Z\kern-.4em Z}}{\hbox{\cmss Z\kern-.4em Z}}
{\lower.9pt\hbox{\cmsss Z\kern-.4em Z}}
{\lower1.2pt\hbox{\cmsss Z\kern-.4em Z}}\else{\cmss Z\kern-.4em Z}\fi}
\def\IGa{\relax\hbox{${\rm I}\kern-.18em\Gamma$}}
\def\IPi{\relax\hbox{${\rm I}\kern-.18em\Pi$}}
\def\ITh{\relax\hbox{$\inbar\kern-.3em\Theta$}}
\def\IOm{\relax\hbox{$\inbar\kern-3.00pt\Omega$}}

\def\np{Nucl. Phys. }
\def\pl{Phys. Lett. }

\def \CA{{\cal A}}

\def \CA{{\cal A}}

\def \l{\ell}

\def \sinh{{\rm sinh}}
\def \cosh{{\rm cosh}}
\def\CD {{\cal D}}

\def\p {\partial}
\def\CS {{\cal S}}
\font\cmss=cmss10 \font\cmsss=cmss10 at 7pt
\def\IZ{\relax\ifmmode\mathchoice
{\hbox{\cmss Z\kern-.4em Z}}{\hbox{\cmss Z\kern-.4em Z}}
{\lower.9pt\hbox{\cmsss Z\kern-.4em Z}}
{\lower1.2pt\hbox{\cmsss Z\kern-.4em Z}}\else{\cmss Z\kern-.4em Z}\fi}

\def\l {\ell }

\def\p {\partial}
\def\CS {{\cal S}}

\def\CZ {{\cal Z}}
\def\CA {{\cal A}}
\def\CG {{\cal G}}

\def\CD {{\cal D}}
\def\pl{{\it Phys. Lett. } }
\def\np{{\it Nucl. Phys.} }

\def\R{\relax{\rm I\kern-.18em R}}
\font\cmss=cmss10 \font\cmsss=cmss10 at 7pt
\def\Z{\relax\ifmmode\mathchoice
{\hbox{\cmss Z\kern-.4em Z}}{\hbox{\cmss Z\kern-.4em Z}}
{\lower.9pt\hbox{\cmsss Z\kern-.4em Z}}
{\lower1.2pt\hbox{\cmsss Z\kern-.4em Z}}\else{\cmss Z\kern-.4em Z}\fi}
\lref\Iml{I.K. Kostov, \pl B 266 (1991) 42}
\lref\Icar{I.Kostov, ``Strings embedded in Dynkin diagrams'',  Lecture
given at the Cargese meeting, Saclay preprint
SPhT/90-133.}
\lref\bkz{E. Br\'ezin, V. Kazakov, and Al. B. Zamolodchikov,
\np 338 (1990) 673}
\lref\Iade{I. Kostov, \np  326 (1989)583.}
\lref\Inonr{I. Kostov, \pl 266 (1991) 317.}
\lref\df{V.Dotsenko and V. Fateev, \np 240 (1984) 312}
\lref\Imult{I. Kostov,
\pl 266 (1991) 42.}
\lref\ajm{J. Ambjorn, J. Jurkiewicz and Yu. Makeenko, \pl 251B (1990)517}
\lref\mat{V. Kazakov, \pl 150 (1985) 282;
F. David, \np 257 (1985) 45;
V. Kazakov, I. Kostov and A.A. Migdal, \pl 157 (1985),295;
J. Ambjorn, B. Durhuus, and J. Fr\"ohlich, \np 257 (1985) 433}
\lref\brkz{E. Br\'ezin and V. Kazakov, \pl 236 (1990) 144.}
\lref\Ion{I. Kostov, {\it Mod. Phys. Lett.} A4 (1989) 217}
\lref\dglsh{M. Douglas and S. Shenker, \np B 335 (1990) 635.}
\lref\grmg{D. Gross and A. Migdal, {\it Phys. Rev. Lett.} 64 (1990) 127.}
\lref\mike{M. Douglas, \pl 238 (1990) 176.}
\lref\pol{A. Polyakov, \pl 103  (1981) 207, 211.}
\lref\kpz{A. Polyakov, \pl 103B (1981) 253;
V. Knizhnik, A. Polyakov and A. Zamolodchikov,{\it Mod. Phys. Lett.}
A3 (1988) 819.}
\lref\ddk{F. David,{\it Mod. Phys. Lett.} A3 (1988) 1651; J. Distler and
H. Kawai, \np 321 (1989) 509}
\lref\bk{M. Bershadski, I. Klebanov,\np 360 (1991) 559. }
\lref\mss{G.Moore, N.Seiberg and M. Staudacher, \np B 362 (1991) 665.}
\lref\polci{J. Polchinski, \np 346 (1990) 253 }
\lref\stau{M. Staudacher, \np 336 (1990) 349.}
\lref\gm{D. Gross and A. Migdal, \np340 (1990) 333}
\lref\brkz{E. Br\'ezin and V. Kazakov, \pl  236 (1990) 144}
\lref\dglsh{M. Douglas and S. Shenker, \np 335 (1990) 635}
\lref\grmg{D. Gross and A. Migdal, {\it Phys. Rev. Lett.} 64 (1990) 127}
\lref\mike{M. Douglas, \pl 238 (1990) 176}
\lref\pol{A. Polyakov, \pl 103 (1981) 207, 211}
\lref\moore{G. Moore, \np 368 (1992) 557}
\lref\gr{I.S. Gradshteyn and I.M. Ryzhik, {\it Table of Integrals, Series
and Products}, Academic Press, 1980.}
\lref\gkl{ D. Gross and I.Klebanov, \np 344 (1990) 475}
\lref\irrationalivan{I.K. Kostov, \pl 266 B (1991) 317}
\lref\Imult{I.K. Kostov, \pl 266B (1991) 42}
\lref\ivanlat{I.K. Kostov,  \np B 376 (1992) 539}
\lref\Iade{I.K.Kostov, \np 326 (1989) 583}
\lref\dfkt{P. Di Francesco and D. Kutasov, \pl 261 B (1991) 385;
P. DiFrancesco and D. Kutasov, \np 375 (1992) 119}
\lref\kusei{D. Kutasov and N. Seiberg, \np 358 (1991) 600}
\lref\bk{M. Bershadski  and I. Klebanov, \np 360 (1991) 559}
\lref\coul{ B. Nienhuis, {\it in Phase transitions
and critical Phenomena} , Vol. 11, ed. C.C. Domb
and J.L. Lebowitz (Academic Press, New York, 1987) ch. 1;
 P. Di Francesco, H. Saleur and J.-B. Zuber,
{\it J. Stat. Phys.} 49 (1987) 57 }
\lref\msei{G. Moore and N. Seiberg, {\it Int. Journ. Mod. Phys.}
A7 (1992) 2601}
\lref\kami{V. Kazakov and A. Migdal, \np 311 (1989) 171}
\lref\daje{S. R. Das and A. Jevicki, {\it Mod. Phys. Lett. }A5 (1990) 1639}
\lref\mqma{E. Br\'ezin, V. Kazakov and Al.B. Zamolodchikov,
\np 338 (1990) 673; D.J. Gross and N. Miljkovic, \pl 238 B (1990) 217;
G. Parisi, \pl 238 B (1990) 209; P. Ginsparg and J. Zinn-Justin,
\pl 240 B (1990) 333}
\lref\mqmb{D.J. Gross and I.R. Klebanov, \np 344 (1990) 475;
\np 352 (1991) 671; \np 354 (1991) 459; \np 359 (1991) 3}
\lref\mqmc{For a review containing further references see:
I.R. Klebanov,``String Theory in Two Dimensions'',
1991 Trieste lectures, Princeton preprint PUPT-1271}
\lref\mqmd{Latest developments are described and referenced in:
R. Dijkgraaf, G.Moore and R.Plesser, ``The Partition Function of 2D
String Theory'', Princeton-Yale preprint IASSNS-HEP-92/48,
YCPT-P22-92}
\lref\IMat{I. Kostov and
 M. Staudacher, Multicritical phases
of the $O(n)$ model on a random lattice, Saclay preprint SPhT/92/025,
to appear in \np }.
\lref\ivmm{I.K. Kostov, ``Gauge invariant matrix model for A-D-E closed
strings'', to be submitted to \pl }

\Title{}
{\vbox{\centerline
{Propagation of Particles and Strings}
\vskip2pt
\centerline{
in One-Dimensional Universe
 }}}

\vskip6pt

\centerline{Ivan K. Kostov \footnote{$ ^\ast $}{on leave of absence
from the Institute for Nuclear Research and Nuclear Energy,
Boulevard Tsarigradsko shosse 72, BG-1784 Sofia, Bulgaria}}

\centerline{{\it Service de Physique Th\'eorique
\footnote{$ ^\dagger$}{Laboratoire de la Direction des Sciences
de la Mati\`ere du Comissariat \`a l'Energie Atomique} de Saclay
CE-Saclay, F-91191 Gif-Sur-Yvette, France}}

\vskip .3in

\baselineskip10pt{
We review the basics of the dynamics of closed strings moving along the
infinite discretized line \Z. The string excitations are described by a field
$\varphi_x(\tau)$ where $x\in \Z$ is the position of the string in the
embedding space and $\tau$ is a semi-infinite ``euclidean time'' parameter
related to the longitudinal mode of the string. Interactions due to splitting
and joining of closed strings are taken into account by a {\it local} potential
and occur  only along the edge  $\tau =0$ of the semi-plane $(x, \tau)$.
}

\vskip 1cm
\leftline{Lecture delivered at the   4th Hellenic School on
Elementary}
\leftline{ Particle Physics, Corfu, 2-20 September 1992}
\bigskip
\rightline{SPhT/93-038}
\Date{April 1993 }

\baselineskip=16pt plus 2pt minus 2pt

\leftline{ \bf 1. Introduction}
\smallskip
The standard quantum field theory operates with point-like excitations
(particles). From a certain point of view it can be considered as
statistical mechanics of interacting random walks. However, there are
theories (perhaps the most interesting ones) for which this scheme
is too restrictive. Instead of propagation of particles we have to deal with
the random motion of extended objects (closed and open strings).
For example, it is believed that  the multicolour (large $N$) limit of Quantum
Chromodynamics  \ref\thoo{G. 't Hooft, \np B 72 (1974) 461} is equivalent
 to some kind of string theory. The world surface of
 of the chromodynamical string then would appear as the result of
condensation of high-order planar diagrams.

This was the main motivation for  some physicists (including the author)
to try to understand the simplest theory of strings whose path integral
is given by the sum over all surfaces swept by the string weighed by
exp[$-$ cosmological constant $\times$ area]. This model can be considered
 as the two-dimensional
generalization of the problem of random walks.
The model is correctly defined only if the dimension $D$ of the
target space is $\le 1$; in higher dimensions the appearance of a tachyon
with $ ({\rm mass})^{2} \sim 1-D$ leads to instability of the vacuum
and the world sheet of the string degenerates into a ramified one-dimensional
object.

The $D=1$ string has been studied carefully in the last time and in sertain
sense  could be considered solved.
Besides the standard quantization due to Polyakov \kpz ,  there are
two gauge-invariant approaches inspired by the discretization of the
world sheet through planar graphs.
The first approach  exploits the possibility to formulate the
$D=1$ bosonic string theory
as the  quantum mechanics of a large $N$ hermitian
matrix \kami. This model  can be rewritten as an intriguingly simple
system of free fermions which upon bosonization turns into
a rather unusual field theory with only one interaction
vertex \daje. The model has been exhaustively solved in
references \ref\kone{I. Kostov, \pl B 215 (1988) 499}
\mqma, \mqmb, \moore, \mqmc, \mqmd
(and references therein). The method of solution
appears however rather removed from the traditional image of a
 string field theory which incorporates  factorization, sewing and
infinitely many interactions. It is precisely these concepts
which appear naturally in the second (loop gas) approach where the
embedding space is also discretized \Iade , \irrationalivan, \Imult , \ivanlat.

In this lecture we will try to explain the logic of the loop gas approach
to the string field theory restricting ourselves to the simplest case when
the target space of the string is the discretized line $\Z$.
\foot{
The techniques apply
equally  to the case when the target space is an ADE or ${\rm \hat A \hat D
 \hat E}$
  Dynkin diagram \Iade . In fact, one of the strenghts of the loop gas
approach is a {\it unified} description of all $D\leq 1$
noncritical strings.
}
Our construction will exploit  the fact that in the loop gas model
 the two components of the dynamics of the
world surface - the  evolution of the intrinsic geometry
 (splitting and
joining of strings) and the  motion in the embedding space - can be
completely  separated.
As a worm-up exersice we will consider first the rather
trivial problem of random  walks in $\Z$.
\smallskip
\leftline{ {\bf 2. Random walks in \Z}}
\smallskip
The Euclidean transition amplitude $G(x,x')$ for a quantum relativistic
particle is given by the sum over all paths connecting the points $x,
x' \in \Z$, each path entering with weight exp[$-$ mass $\times$ length].
A path $\Gamma_{xx'}$ in $\Z$ is by definition a map of the
 proper-time interval
$\gamma _{\l}=[0,\l]$ in $\Z$. Each point $\xi \in \gamma_{\l}$ has an
integer co-ordinate (higth) $x(\xi)\in \Z$.
The two-point amplitude then reads
\eqn\i{G(x,x')= \int_{0}^{\infty} d\l
\sum _{\Gamma_{xx'}}e^{-m_{0}\l}}
where $m_{0}$ is the bare mass and $\l $ is the
length (= proper time) of the evolution line of the particle.

The sum on the r.h.s. of \i\ can be thought of as the partition function
of an SOS model defined on the world line of the quantum particle.
This sum can be performed as follows. First we notice that the world line
embedded in $\Z$ can be divided into segments with constant higth
(we assume
that the higth can jump only by $\pm 1$). 
Then by separating the first segment we write a simple recursion
equation for $G(x,x')$:
\eqn\ii{G(x,x')={1 \over m_{0}}(G(x+1,x')+G(x-1,x')+\delta_{x,x'})}
which is a discretized version of the Klein-Gordon equation.

Considering $x$ as a continuum variable and introducing the parametrization
$m_0=2\cosh (\pi E)$ we write \ii\ in the following operator form
\eqn\iii{(e^{\pi E}+e^{-\pi E}-e^{\p_x}-e^{-\p_x})G(x,x') = \delta_{x,x'}}
This equation diagonalizes in the Fourier space (which is periodic
with period 2)
\eqn\iv{G(p)={1 \over 2 \cosh \pi E - 2\cos \pi p}
}
Finally, let us mention that the sum over all closed paths in $\Z$ is
formally obtained as the vacuum energy
of the following simple field theory
\eqn\kas{
Z=\int \prod _{x\in \Z}d\phi_x \ \exp \big(\sum_{x}
\phi_x  \big(  \cosh \p_x - \cosh (\pi E)
\big) \phi_x \big)}
and the transition amplitude \i\ is given by the two-point correlator
\eqn\vcv{G(x,x')= \langle \phi_x \phi_{x'} \rangle . }
\smallskip
\leftline{ \bf 3. Random surfaces in \Z}
\smallskip
It turns out that the above analysis can be generalized
to  the case of  random motion of closed strings  immersed in $\Z$.
The   closed-string states
 are determined completely by  the intrinsic length $\l$ and the occupation
$x$ in the target space $\Z$. The loop-loop correlator
$G(x, \l ;x',  \l ')$  is defined as
the sum over all cylindrical surfaces $\CS_{xx'}$ immersed in $\Z$
  and having the two loops as boundaries
\eqn\surs{G(x,\l;x',\l ')=\int _{0}^{\infty}dA
 \sum_{\CS_{xx'}} e^{-\Lambda A }}
where the parameter $\Lambda$ coupled to the area $A$ of the
surface $\CS_{xx'}$ is usually called  ``cosmological constant''.
 The sum contains an
integral over all world-sheet geometries and all possible ways to map given
 world sheet in $\Z$.  As in the case of the random particle, the
world  sheet can be divided into domains of constant height $x$
which are separated by domain walls (fig. 1).
\vskip 5cm
\centerline{{\bf Fig.1.}  A piece of a surface immersed in $\Z$:
 domains and domain walls}
\smallskip
Our further strategy will be based on the trivial observation that
 the decomposition of the world sheet of
the embedded surface into domains of constant height generates a decomposition
of the integration measure in the space of  surfaces into a sum over Feynman
graphs.
In the problem of random surfaces
the things are technically  more complicated: a domain on the world sheet
is characterized not only by the lengths of its boundaries
$\l_1,\l_2,...$, but also by its intrinsic
 geometry (curvature) which can exhibit local
fluctuations. The string interaction constant $\kappa$ is coupled to the total
curvature $\chi = 2-2h-n$ where $h$ is the number of handles and
$n$ is the number of boundaries. The Boltzmann weight of a domain
 is given by the sum
over all possible geometries of a surface having $h$ handles and $n$ boundaries
with lengths $\l_1,...,\l_n$. This sum can be evaluated using the
discretization of the configuration space of surfaces {\it via} planar graphs.
We will not go into the details since there exists abundant literature on this
subject
(see, for example,
\ref\car{Proceedings of a NATO Advanced Research Workshop on
Random Surfaces and Quantum Gravity, held May 27 - June 2, 1990,
in Carg\`ese, France, 1991 Plenum Press, New York} and the references therein).
 The general $n$-loop amplitude was calculated
 explicitly in the planar limit $\kappa =0$ in  \ajm\ and \mss .
 In the simplest case when the area of the domain is zero, this result can
be generalized to the case of general topology
using the loop equations for the  gaussian  matrix model \Icar  , \ivanlat
\eqn\oiu{\eqalign{
W_{n}(\l_1,\l_2,...,\l_{n})&=
\sqrt{\l_1 ... \l_n} (\l_1+...+\l_n)^{n-3}e^{-\sqrt{\mu}(\l_1+...+\l_n)+
\kappa^2 (\l_1+...+\l_n)^3}\cr
&= 
 \int_{- \infty}^{-\sqrt{\mu}} dz \  \p_z^{n-2}
 e^{\kappa^2 \p _z ^3}
\ \prod_{k=1}^n \sqrt{\l_k}e^{z\l_k}\cr}}
(Note that this formal expression is divergent for any finite $\kappa$;
this is a well known desease of the bosonic string theory.)

It is possible to construct the path integral of the string using these
simple amplitudes; then the boundaries between domains of constant hight $x$
(i.e., the domain walls)
will be distributed densely on the world sheet and the
role of cosmological constant will be played by the
parameter $\mu$ coupled to the length of the domain walls.
 The two-loop propagator then has the meaning of the partition function for
the  SOS
 model on a cylindrical  random surface
with fluctuating geometry and  constant hights $x$ and $x'$ along its two
 boundaries.

Similarly to the  case of random walks we can construct a `` string field
theory'' for the loop field $\Phi_{x}(\l)$ in which
 the loop-loop
correlator is given by the sum over surfaces \surs .
The partition function of this field theory is given by
 the following functional integral
\eqn\pf{\eqalign{
Z&=\int \CD \Phi \exp^{-\CA [\Phi] } \cr
 \CA &={1 \over 2} \sum _x \int _{0}^{\infty}{d\l \over \l}
 \Phi_{x}(\l) [e^{\p_x}+e^{-\p_x}]^{-1}
 \Phi_{x}(\l) \cr
& - \sum _{n =1 }^{\infty} {\kappa^{n-2} \over n!}\sum_{x}
\int {d\l_{1}\over \l_1}...{d\l_{n}\over \l_n}
 W_{n}(\l_1,\l_2,...,\l_n)\Phi_{x}(\l_{1})  ...\Phi_{x}(\l_{n})\cr }}

The vertices in the interaction potential represent surfaces
 of various topologies embedded in a single point $x \in \Z$.
 The  kinetic term of the action leads to a ``propagator'' which
can be thought of as an infinitesimal cylinder connecting two edges with
co-ordinates $x$ and $x+1$.
The  Feynman graphs constructed in
 this way    represent surfaces embedded in $\Z$.
 The weight of a graph with $h$ handles and $n$ boundaries depends
on the string interaction constant through  a factor
$\kappa^{2-2h-n}$.

This diagram technique contains  a tadpole vertex $W_1$ (disk)
and a two-leg vertex $W_2$ (cylinder). We will bring it, by
a re-definition of the field, to a standard form where the interaction
starts with a $\phi^3$-term.

In order to elliminate the tadpole we shift
 the string field by its expectation
value
\eqn\vev{\Phi_x(\l)=\langle \Phi_x(\l)\rangle _{\kappa =0} +\phi_x (\l)}
The classical string field is equal to the sum over all planar surfaces
spanning a loop with length $\l$. It can be evaluated as the solution of
the saddle-point equation for the theory \pf\
 which is equivalent to the  loop equation
in the planar limit. Its explicit form reads \Icar
\eqn\kzero{\eqalign{
\langle \Phi_x(\l ) \rangle_{\kappa =0} &=
\int^{\infty}_{\sqrt{\mu}} dz [(z+\sqrt{z^2-\mu})^g-(z-\sqrt{z^2-\mu})^g]
e^{-z\l}\cr
&= \mu ^{g/2}
{2g \over \l} K_{g}(\sqrt{\mu} \l), \ \ \ g=1 }
}%
where $K_g$ is a modified Bessel function
\eqn\bssbl{K_{g}(\sqrt{\mu}\l)= \int _{0}^{\infty}
 d\tau e^{-\sqrt{\mu} \l \cosh \tau}
\cosh (g \tau)}
The space of possible backgrounds is parametrized by the positive constant $g$;
the vacuum in our case is translationary invariant which corresponds to the
choice $g=1$.
The  classical solution  with $g\ne 1$ describes a
string beckground with  non-zero momentum $p_0=g-1$.
This is the case, for example,  when
the target is restricted to a chain of $m-1$ points (i.e., the ${\rm A}_{m-1}$
Dynkin diagram)  and the
 translational invariance is lost. In this case $1-g=1/m$.
In particular, for  $m=2$, the classical field  \kzero\ coincides
with  the tadpole vertex $W_1$ given by eq.  \oiu .

The linear change \vev\ leads to new vertices $w_n(\l_1,...,\l_n)$
but they have the same
structure as the original ones \oiu .
The first ones are \ivanlat
\eqn\xes{w_2(\l,\l')|_{\kappa =0} =
\sqrt{\l}{e^{-\sqrt{\mu}(\l +\l ')} \over \l +\l '} \sqrt{\l '}}
\eqn\polk{w_3(\l _1,\l _2, \l_3)|_{\kappa =0}= \mu^{-(1/2-g)/2}
\prod_{k=1}^3 \sqrt{ \l_k} \ e^{-\sqrt{\mu}\l_k}
}
\eqn\jiij{w_4(\l _1,\l _2, \l_3,\l_4)_{\kappa =0}= \mu^{(1/2 -g)}
\prod_{k=1}^4\sqrt{\l_k} \ e^{-\sqrt{\mu}\l_k} [\sum_{k=1}^4
\l_k +{1 \over 2M}(g^2-{1\over 4})]}
Note that the two-loop vertex does not change after the shift
\vev . Indeed,
the new vertices make sense of
loop amplitudes of a special one-matrix model \ivanlat\  and it is well
known that the two-loop correlator in all matrix models is universal.
The planar ($\kappa =0$) vertices $w_n$  can be evaluated according to the
general formula for the planar  loop amplitudes in the one-matrix model
\ajm \mss
\eqn\plmn{
w_n(\l_1,...,\l_n)= \Big( {d \over d t } \Big)^{n-3}
\prod_{k=1}^n \sqrt{\l_k}e^{-f\l_k} |_{t=0}
}
where the function $f(M,t)$ is determined by the following
``string  equation''   \ivanlat
\eqn\ffsf{ t = M^{g-1/2}(M-f)\sum_{k=0}^{\infty}{(1/2+g)_k(1/2-g)_k
\over k!(k+1)!}\Big({M-f \over 2M} \Big)^k}
so that $f(M,0)=M$.
 We used the standard notation $(a)_k = a(a+1)...(a+k-1)$.
 Note that the new vertices coincide with the old ones
when $g=1/2$.

The next step is to evaluate the propagator in this string field theory
which is given by the sum \surs\ of cylindrical surfaces immersed in $\Z$.
For this purpose we need to invert the quadratic form in the
 gaussian part of the action. The latter is given by the sum of the
 kinetic term in \pf\ and  the two-loop vertex $w_2$.
This will be the subject  the next section.
\smallskip
\leftline{ \bf 4. A quantum field theory on the semi-plane}
\smallskip
Below we will see that, in specially chosen variables, the string field
theory can be driven in to a more usual form of a field theory with
local interaction, defined on the semi-plane. The interaction
will be concentrated along the edge of the semi-plane.
This field theory reminds the  effective  string field theory
of Polchinski \polci\ in which the interaction increases exponentially
along the negative time direction so that the theory is effectively
confined to a semi-plane.

The diagonalization of
 the gaussian  ($\kappa =0$) part of the action
is done by means of  the formula
 \ref\grrz{I.Gradshtein and I. Ryzhik, Table of
Integrals, Series, and Products, Academic press, 1965}
\eqn\jhg{
\sqrt{\l}{e^{-\sqrt{\mu}(\l +\l ')} \over \l +\l '} \sqrt{\l '}=
{2 \over \pi} \int _{0}^{\infty} dE \ E \tanh (\pi E)
K_{iE}(\sqrt{\mu}\l)K_{iE}(\sqrt{\mu}\l')}
We introduce, following \mss ,  a complete set of
delta-function normalized eigenstates \mss
\eqn\ytro{\langle x, \l |p, E\rangle ={2\over \sqrt{ \pi}}
\sqrt{E \sinh(\pi E)}\  e^{i\pi p x} K_{iE}(\sqrt{\mu}\l) , \ \ \
 E>0, \ \  -1<p\le 1 }
\eqn\egss{ \sum_{x} \int _{0}^{\infty}{d\l \over \l}
 \langle p, E|x, \l \rangle \langle x, \l |p', E' \rangle
= \delta (E, E') \delta (p,p')}
Then we write the functional integral in terms of the amplitude
$\varphi (p,E)$ which we normalize as follows
 \eqn\hyh{\varphi (p,E)={1 \over \pi} {\sqrt{ \pi E \sinh ( \pi E)}
\over  \cosh (\pi E)}
\sum_x \int {d\l \over \l}
\langle p,E| x, \l \rangle \phi_x ( \l)
}
After that the free part of the action takes the form
\eqn\cdf{
 \CA_{\rm free} ={1 \over 2}\int_0^{\infty}dE \int_{-1}^1 dp\
\varphi(p,E)
\CG^{-1}(p,E)
\varphi (p, E)}
\eqn\xzsz{\CG (p,E)  = { \pi E \sinh (\pi E)
\cos (\pi p) \over 2 \cosh (\pi E)[\cosh (\pi E) -  \cos (\pi p)]}}

The general term in the interaction part of the action
can be transformed to the $E$ space
using the formula \grrz
\eqn\ddg{
\int _{0}^{\infty} {d\l \over \l}\ \sqrt{ \l} e^{\sqrt{\mu}\l}
 \l ^k K_{iE}(\sqrt{\mu} \l) =(2\sqrt{\mu})^{-k-1/2}
{(1/2+iE)_k (1/2 -iE)_k \over k!}{\pi^{3/2} \over \cosh (\pi E)}
}
It follows from the general form of the interaction
\oiu\ that the planar ($\kappa =0$)
  $n$-point vertex  in the $E$ space will be an odd
polynomial of  $E_1,...,E_n$ of total  degree $2(n-3)+6h$ where
$h$ is the number of handles of the corresponding elementary surface.
The  lowest  vertices \polk , \jiij\  in the momentum space are
\eqn\aasa{ w_3(E_1,E_2,E_3)=1}
\eqn\lkl{w_4(E_1, E_2,E_3, E_4)=
(g^2-1/4)+\sum_{k=1}^4 (1/4 +E_k^2)}

Let us introduce a time  variable $\tau$ dual to the quantum number $E$
and reformulate  the functional
integral in terms of the field
\eqn\vfv{\varphi_x (\tau )=
\int_{-1}^{1} dp \int_{0}^{\infty} dE e^{i\pi p x}
 \cos ( E\tau ) \varphi (E,p), \ \ \ \ \ \ \tau \ge 0 }

Then  the gaussian part of the action takes
 the form
\eqn\aax{\CA_{\rm free}=\sum_x\int_0^{\infty} d \tau
\varphi_x (\tau)
{
[\cos (\pi \p_{\tau})-\cosh \p_x ]\varphi_x (\tau)
\cos (\pi \p_{\tau})
 \over 2\pi \p_{\tau} \sin (\pi \p_{\tau}) \cosh (\pi \p_x)}
}
The interaction is concentrated along the wall $\tau =0$ in the $(x,\tau)$
space. For example, the two lowest vertices   \aasa , \lkl\ are generated by
the potential
\eqn\ltr{ \CA_{\rm int} =  \sum_x \Big(\kappa
{[ \varphi_x(0)]^3 \over 3!}
 +\kappa^2
\big( g^2 -{1\over 4}\big){[ \varphi_x(0)]^4 \over 4! }
  +\kappa^2  {[({1 \over 4}-\p_{\tau}^2) \varphi_x(0)] \over 1! }
{[\varphi_x(0)]^3 \over 3! }\Big)}
The field $\varphi_x(\tau)$ is related to the original field
$\phi_x(\l)$ by the relations
\eqn\plok{\varphi_x(\tau)={1 \over \pi}
{\p_{\tau}\sin(\pi \p_{\tau}) \over \cos (\p_{\tau})}
\int_{0}^{\infty} {d\l \over \l} e^{-\sqrt{\mu}\l \cosh \tau}
\phi_x(\l)}
\eqn\plko{\phi_x(\l)= \int _0^{\infty}d\tau
\  e^{-\sqrt{\mu}\l \cosh \tau}
 \cos (\p_{\tau}) \varphi_x(\tau)}
which  follow from the integral representation of the Bessel function \bssbl .
The  transformation between  the eigenstates in the
 $\l$ and $\tau$ spaces has been first discussed
in \msei .
The field $\varphi_x(\tau)$  is  restricted to the
semi-plane $\tau \ge 0$,  but it can be considered as defined on the
whole  plane $(x,\tau)$ if the
symmetry  $ \varphi_x(-\tau)=\varphi_x(\tau)$ is imposed.

Finally let us make the following remark concerning the propagator of
our string field theory.
It is natural to expect that upon diagonalization it decomposes into a
 sum of random-walk
propagators with different masses. On the other hand, the action \aax\
does not seem to meet this requirement.
The reason is the following.
The propagator \xzsz\ can be decomposed
as a sum of two terms
\eqn\ppppa{
\CG(E,p) = G(E,p)-G(E,1/2)}
with
\eqn\popl{G(E,p)={\p E \sinh (\pi E) \over
2  \cosh (\pi E)-
2\cos (\pi p)]}= {E\over 2} {d \over dE}\log (\cosh (\pi E)-\cos (\pi p))
}
The function \popl\ is, up to a numerical factor, equal to the propagator of
a random particle moving in $\Z$.

This form of the propagator admits simple explanation. The target space
of a string theory  without
embedding  consists of a single point; the corresponding momentum space
 contains a single
momentum $p=1/2$.
The corresponding string field theory has vanishing propagator and its loop
amplitudes are the vertices \plmn .
The full string field theory can be considered as a perturbation of the
$p=1/2$ string field theory.

  Of course, it is possible to absorbe, by
 re-definition of the vertices, the $p=1/2$ term into the interaction.
This is however not advisible since then the $E$-integration would
produce  singularities. These singularities
 are neatly cancelled for the difference \ppppa .

Note also that the Fourier image of the discrete propagator \popl\
is obtained from the two-point correlator of a two-dimensional Euclidean
 particle
by ``periodization'' with respect to translations $p \to p+2$
\eqn\id{
G(E,p)
=\sum_{n=-\infty}^{\infty}
{E^2\over E^2 + (p+2 n)^2}
}

\smallskip
\leftline { \bf 4. Loops}
\smallskip
Let us evaluate the  simplest loop diagram
 giving  the partition
 function $
\CZ_{{\rm torus}}$
of  the noninteracting closed
string.

It is convenient first to perform the calculation for  fixed
 momentum $p$ running along the loop.  The corresponding piece of the
partition function consists of two terms
\eqn\pfpf{
\CZ (p)= \CZ(1/2)- \int _{a} ^{\infty} {d\l \over \l}\int_0^{\infty}dE
|\langle \l |E \rangle |^{2}  \log \Big({
 \cosh  (\pi E) -\cos  (\pi p) \over \cosh  (\pi E) - \cos (\pi 1/2)} \Big)}
The contribution of the surfaces with momentum $p=1/2$ is equal to the
$\kappa =0$ term in the expansion of the vacuum energy of the effective
one-matrix model generating the amplitudes \plmn .
 A simple calculation (see, for example Appendix C of \grmg )
yields
\eqn\fser{\CZ(1/2)={g-1/2 \over 24}\log (a\sqrt{\mu})}
The  integral over the length $\l$ in  \pfpf\ yields a factor $\log [1/(a
\sqrt{\mu})]$
where $a$ is a cut-off parameter (this is the diagonal
value of the kernel of the regularized identity operator in the
$E$-space) and the $E$-integral equals
\eqn\immnb{ \int_{0}^{\infty}{dE \over \pi}
\log \Big( {\cosh (\pi E) - \cos (\pi p) \over \cosh(\pi E)}\Big)
= {1 \over 2}(|p|-{1 \over 2})(|p|-{3\over 2}), \ \ \ |p|\le 1}
Inserting this \fser\ and \immnb\ in \pfpf\ we find
\eqn\ppff{ \CZ(p) =
{1  \over 24}
 \big[ g-1/2 + 6(1/2 -|p|)^{2}\big] \log (a\sqrt{\mu})}

This result can be applied to discrete target spaces with various
geometries since the target space is characterized
completely by  the spectrum of allowed momenta
$p$.
Consider, for example, the string compactified on the circle with $2m$
points $\Z_{2m}$.
Summing over all allowed momenta
$p=k/m;\  k=0,\pm 1,...,\pm(m-1),m$, we find  (for the translationally
invariant background $g=1$)
\eqn\uyol{\CZ_{{\rm torus}}
     =
 \Big( {m^{2}+2 \over 24m } + 2m \  {g-1/2 \over 24} \Big) \log (aM)
={1 \over 12}
 \Big(
{m+{1 \over m}
 \Big)\log (aM)}}
This is precisely the result obtained  the partition function of
the string embedded in a continuum
compact target space \gkl \ref\torr{P. Di Francesco and D. Kutasov,
\np B 375 (1992) 119}.
\smallskip
  \leftline{ \bf Concluding remarks}
\smallskip
The most attractive feature of the  loop gas approach
is the possibility to construct a simple diagram technique for the loop
amplitudes. This is possible because the so-called ``special states''
corresponding to momentum-independent poles, are  inobservable in the
string theory with discrete target space. A rather dramatic consequence
is the {\it factorization} of the interactions which may be traced back to
the fact that the string interaction takes place at a single point in the
$x$-space.  Nevertheless, the string theory with target space $\Z$
seems to be identical to the string theory with
continuum embedding space $\R$, if the right observables are compared.
It has been checked \ref\ksks{I. Kostov and M. Staudacher, ``Strings
in discrete and continuum target spaces: a comparison'', preprint
RU-92-21 and SPhT/92-092, to be published in \pl B} that the $n$- loop
amplitudes in $x$ space in both theories coincide  up to $n=4$.
(Actually, since the four loop amplitude in the $\R$-string
is not yet calculated,  the comparison for $n=4$ was restricted to the
case of  infinitesimal
loops.)

Another remark is the the striking similarity of the loop-gas diagram
technique with the  construction of the interacting string theory
by B. Zweibach \ref\bz{B. Zweibach, \np B 390 (1993) 33} in which
the elementary vertices in the nonpolynomial action are themselves
loop amplitudes. A consistent choice of string vertices and propagator
give rise to one particular way of decomposing the moduli space
of Riemann surfaces. The possibility of choosing different decompositions
of moduli space is related to a special symmetry of the string field theory.
Such a symmetry might exist in the loop-gas construction as well.

Finally let us mention that the loop gas approach allows to
equally construct the open string field theory in the space $\Z$
\ref\last{V. Kazakov and I. Kostov, \np B 386 (1992) 520}.

\listrefs

\bye